\def\marker#1{#1}
\def\markers#1{#1}
\def\mmarker#1{$#1$}
\def\mathmarker#1{#1}

\documentstyle[11pt,epsfig]{article}
\makeatletter
\def\lesssim{\mathrel{\mathpalette\gl@align<}}
\def\gtrsim{\mathrel{\mathpalette\gl@align>}}
\def\gl@align#1#2{\lower.6ex\vbox{\baselineskip\z@skip\lineskip\z@
    \ialign{$\m@th#1\hfil##\hfil$\crcr#2\crcr\sim\crcr}}}
\makeatother

\newcommand{\dfrac}[2]{\frac{\strut \displaystyle{#1}}{\displaystyle{#2}}}

\catcode `\@=11
\def\@cite#1#2{#1\if@tempswa , #2\fi}
\catcode `\@=12
\newcommand{\upcite}[1]{${}^{\mbox{\scriptsize \cite{#1}}}$}
\newcommand{\rcite}[2]{${}^{\mbox{\scriptsize \cite{#1}--\cite{#2}}}$}

\begin{document}
\bibliographystyle{prsty}
%
\vspace*{50mm}
\noindent
{\bf MAGNETISM AND ELECTRONIC STATES OF SYSTEMS \\
WITH STRONG HUND COUPLING\footnote{%
To appear in proceedings of the workshop ``Physics of Manganites'',
Michigan State University, July 26--29, 1998, eds.\ by T.A.\ Kaplan
and S.D.\ Mahanti, Plenum Publishing Corporation.
}}

\vspace{15mm}
\noindent
\hspace{25mm}K.\ Kubo,$^{1}$ D.\ M.\ Edwards,$^2$ A.\ C.\ {M.}\
Green,$^2$
 T.\ Momoi$^1$ and H.\ Sakamoto$^1$\\

\vspace{5mm}\noindent
\hspace{25mm}
$^1$Institute of Physics, University of Tsukuba, 
Tsukuba, 

\noindent \hspace{26.7mm}
Ibaraki 305-8571, Japan

\noindent \hspace{25mm}
$^2$Department of Mathematics, Imperial College, London SW7 2BZ, 

\noindent \hspace{26.7mm}
UK

%
%
\vspace{15mm}\noindent
{\bf INTRODUCTION}

\vspace{5mm}
\markers{The fascinating} physics of \markers{the}
perovskite manganites\upcite{Ref1} is governed
by electrons which 
hop among or are localized on Mn$^{3+}$ and Mn$^{4+}$ ions.
 It was pointed out already in 
1951 by Zener\upcite{Zener}
that the strong Hund coupling between $e_{\rm g}$ and $t_{\rm 2g}$
electrons is 
essential in understanding 
the ferromagnetism\upcite{Jonker-VS}
 caused by substitution of triply valent La with doubly valent
 elements such as 
Ca, Sr or Ba in LaMnO$_3$.
The two-fold degeneracy of $e_{\rm g}$ orbitals causes other
important effects in the 
physics of manganites. The existence of the orbital degrees of freedom
leads to  
orbital ordering in the insulating phases.\upcite{Orbital O.} Various
charge and orbital ordered states are 
experimentally observed.\upcite{Charge O.} 
Another important effect is the Jahn-Teller distortion, which causes 
strong correlations between lattice distortions and orbital and
magnetic interactions.\rcite{Goodenough-WAM}{Zhao-CKM}
All these effects are entangled and
\markers{give} rise to \markers{the} rich phase diagrams
and peculiar transport phenomena
of \markers{the}
manganites. Though it might be necessary to take into account
all of these effects at \marker{the} same time in order to explain experimental
results quantitatively, it is useful to extract 
one of these important factors and study its effect in detail, in
order to understand the physics of manganites in a profound way.
In this paper we aim to clarify the effects of the Hund coupling by
employing simplified models. 
First we study the ground state phase diagram of a doubly degenerate
Hubbard model. 
Our main concern is 
the effectiveness of the Hund coupling on the
ferromagnetism. 
We compare the result in one dimension with that in infinite
dimensions, and examine common features and differences between them.
The second model we study is a simple ferromagnetic Kondo lattice 
model or double exchange model. 
We investigate one-particle states in this model using a single-site
approximation and calculate 
the electrical resistivity. We treat the 
localized spins as quantum mechanical ones 
and study the quantum effects on the electronic 
states. 

\vspace{15mm}\noindent 
{\bf Doubly Degenerate Hubbard Model}

\vspace{5mm}
We consider in this section a doubly degenerate Hubbard model described by 
the Hamiltonian
 \begin{eqnarray}
   \label{2-band}
   H&=&-t\sum_{m=1,2 \atop \sigma=\uparrow,\downarrow}
   \sum_{\langle i,j \rangle\in {\rm N.N.}}
       (c_{im\sigma}^\dagger c_{jm\sigma} + h.c.)
       +U\sum_{i,m} n_{im\uparrow} n_{im\downarrow}
       +U^\prime \sum_{i,\sigma,\sigma^\prime} 
                   n_{i1\sigma} n_{i2\sigma^\prime} 
\nonumber \\
& &-J \sum_{i,\sigma,\sigma^\prime}
          c_{i1\sigma}^\dagger c_{i1\sigma^\prime} 
          c_{i2\sigma^\prime}^\dagger c_{i2\sigma}
       -J^\prime \sum_i
         (c_{i1\uparrow}^\dagger c_{i1\downarrow}^\dagger 
          c_{i2\uparrow} c_{i2\downarrow} + h.c.), 
\end{eqnarray}
where $c_{im\sigma}$ ($c^\dagger_{im\sigma}$) denotes the annihilation 
(creation) operator of the electron at site $i$ with orbital $m$(=1 or
2) and spin $\sigma$. The number operators are denoted by
$n_{im\sigma}$.
Hoppings of electrons are assumed to occur between the same orbitals
of nearest neighbor sites. 
In real systems, there are off-diagonal hoppings and also hopping
integrals are anisotropic for $e_{\rm g}$ orbitals, that is, they are 
dependent on the directions of hoppings.
This anisotropy may have an important effect \markers{on} the orbital and 
antiferromagnetic \markers{ordering} in
manganites.\upcite{Ishihara-IM,Shiina-NS}
We take here, however, \marker{the} simplest model which can take account of
the effects of orbital degeneracy and Hund coupling.
The interaction terms in eq.\ (\ref{2-band}) originate from the 
Coulomb interaction between electrons at the same site. The last term,
which transfers two electrons on one orbital to the other, is often
neglected. But this term should be properly considered, since it 
enhances local quantum fluctuations and the coefficient $J'$ is 
equal to that of the Hund coupling $J$ if we assume the 
orbital wave functions are real.
\markers{The interaction} parameters satisfy the relation
\begin{eqnarray}
\label{s-relation}
 U=U^\prime+2J
\end{eqnarray}
for $e_{\rm g}$ orbitals, and we assume this relation in the following.

The Hamiltonian (\ref{2-band}) produces various magnetic correlations
and effective ferromagnetic interactions between electrons on
different sites. 
Let us first consider two electrons on one site. The spin-triplet
states, where two electrons occupy different orbitals, are 
stabilized by the Hund coupling and \markers{have} the lowest energy $U'-J$. 
There are three spin-singlet states; 
one with energy $U'+J$
where two electrons 
occupy different orbitals, and the other two with energies 
$U-J'(=U'+J)$ and $U+J'(=U'+3J)$ 
where electrons occupy the same orbital. 
An effective spin interaction between neighboring sites is derived from this 
one-site spectrum in the strong correlation regime ($U'>J\gg t$). 
Let us consider two nearest neighbor sites each of which is occupied
by a single electron in this regime. 
When two electrons \markers{with} parallel spins \markers{sit}
on different orbitals, 
virtual hoppings of the electrons between two sites lower the energy
by $-2t^2 /(U'-J)$. When two electrons have antiparallel spins, the
energy is lowered by $-2t^2 U/(U^2 -J'^2 )$ or $-2t^2 U'/(U'^2 -J^2 )$
depending on whether they are on the same or different orbitals.
\marker{Hence} there \marker{is}
an effective interaction between neighboring sites 
which favors ferromagnetic spin alignment \marker{but}
alternating
alignment of orbital degrees of freedom.\upcite{Roth}
In a system with quarter-filled bands, i.e.\ for 
$n\equiv N_{\rm e}/N=1$, both ferromagnetic long-range order (LRO) 
and alternating orbital order are
expected to coexist in the ground
state.\rcite{Roth}{Inagaki75}
Here $N_{\rm e}$ 
and $N$ denote the total number of electrons and sites, respectively.
The second-order perturbation from the atomic limit ($t=0$)
leads to the following effective Hamiltonian for spin operators 
${\bf S}_i$ and
pseudo-spin operators {\boldmath $\tau$}$_i$\upcite{KugelK,Momoi-K}:
\begin{eqnarray}
  \label{eq:H_eff}
H_{\rm eff}&=& -{t^2} \sum_{\langle i,j \rangle}
  \Biggl[ \dfrac{4U}{U^2-{J^\prime}^2}
  \biggl(\dfrac{1}{4}+\tau_i^z\tau_j^z\biggr)
  \biggl(\dfrac{1}{4}-{\bf S}_i \cdot {\bf S}_j \biggr) \\
&&-\dfrac{2J^\prime}{U^2-{J^\prime}^2}
  (\tau_i^-\tau_j^- + \tau_i^+\tau_j^+)
   \biggl(\dfrac{1}{4}-{\bf S}_i \cdot {\bf S}_j\biggr) \nonumber\\
&&+\dfrac{2U^\prime}{ U^{\prime 2}-J^2 }
  \biggl\{ \dfrac{1}{4}
   - \tau_i^z \tau_j^z
   - 2(\mbox{\boldmath $\tau$}_i \cdot \mbox{\boldmath $\tau$}_j
    - \tau_i^z \tau_j^z)
    \Bigl(\dfrac{1}{4}+{\bf S}_i \cdot {\bf S}_j\Bigr) \biggr\}\nonumber\\
&&+\dfrac{2J}{ U^{\prime 2}-J^2 }
\biggl\{ \tau_i^z \tau_j^z
- \mbox{\boldmath $\tau$}_i \cdot \mbox{\boldmath $\tau$}_j
+ 2\Bigl(\dfrac{1}{4}-\tau_i^z \tau_j^z\Bigr)
\Bigl(\dfrac{1}{4}+ {\bf S}_i \cdot {\bf S}_j\Bigr)
\biggr\} \Biggr].\nonumber
\end{eqnarray}
In strongly correlated systems with the filling $1<n<2$, each lattice
site is either singly- or doubly-occupied, and 
electrons hop from doubly-occupied sites to singly-occupied ones. 
Doubly-occupied sites are almost necessarily in spin-triplet states
due to the Hund coupling and 
the hopping probability is largest between pairs of sites with parallel 
spins. As a result the kinetic 
energy is lowered by ferromagnetic spin correlations. 
This mechanism favoring ferromagnetism is 
quite similar to that in the double exchange model of electrons, where
electrons interacting with localized spins have lower kinetic energy
when spins are aligned parallel.\upcite{Zener,Anderson-H} 
In the following we call this mechanism which favors
ferromagnetism \markers{the}
\markers{``}double exchange mechanism\markers{''} even when we are not
treating localized spins. 
In the case with less-than-\marker{quarter}
filling ($n<1$), the ``double exchange 
mechanism'' may not work for $U'-J\gg t$. Nevertheless the Hund coupling 
may \marker{lead to} ferromagnetism even for
$n<1$, if $t/(U'-J)$ is not too small. The effective
ferromagnetic interaction described by eq. (\ref{eq:H_eff}) 
between nearest neighbor electrons may have a sizable effect \marker{and}
cause metallic ferromagnetism. Van Vleck argued 
 that this mechanism may be operative in realizing
ferromagnetism in Ni.\upcite{VVleck} 

Though \marker{the} mechanism favoring ferromagnetism can be understood
qualitatively as above, it is far from trivial whether 
ferromagnetic long-range order occurs in bulk systems. In the
following, we present a numerical study of the model in one and
infinite dimensions. 

\vspace{5mm}\noindent
{\bf One-Dimensional Model}

\vspace{5mm}
There are rigorous proofs for the ferromagnetic ground state of the
one dimensional model in strong coupling
limits.\rcite{Kubo}{Shen}
These proofs are \marker{valid} in different limits of strong coupling. 
For the strong Hund coupling case ($J\rightarrow \infty$ and
$U\rightarrow \infty$), existence of ferromagnetism is
proved for arbitrary $U'(>0)$ in $1<n<2$,\upcite{Kubo,KusakabeA1} 
and also for $0<n \le 1$ in the special limit $J=U' \rightarrow \infty$ 
and $U\rightarrow \infty$.\upcite{KusakabeA1} Shen obtained a rather
general result that the ground state is fully spin-polarized for any $n$
between 0 and 2 except for 1 if 
$U=\infty$, and $U'(>0)$ and $J=J'(>0)$ are finite.\upcite{Shen} 
(We note that this result cannot be applied naively to our case 
which assumes the relation (\ref{s-relation}).)

So far several numerical studies were done, and 
ferromagnetism was
found for densities near quarter
filling.\rcite{GillS,KusakabeA1,KusakabeA2}{Hirsch}
These studies were done by diagonalizing relatively small systems with
sizes up to 12, and size dependence was not studied yet. 
We thus need to study systems with larger sizes and examine size
effects to obtain conclusive results. 
We note \marker{also}
that most 
previous studies \markers{did} not take into account 
\marker{the} $J'$-term, \marker{and}
\markers{assumed} relations between $U$, $U'$ and $J$
\marker{which differ from} eq.\ (\ref{s-relation}).

We report in the following a study of finite-size chains with up to 16
sites applying the exact diagonalization method as well as the density
matrix renormalization group (DMRG) method. 
We employ open boundary conditions, since the 
periodic boundary condition causes very large size dependence in one
dimension (e.g. even-odd oscillations). Remarkably we found little 
size dependence due to the use of the open boundary conditions. 
Details of this study will be published elsewhere.\upcite{Sakamoto-MK} 

First, we show the ground-state 
phase diagram for $n=1$ in Fig.\ 1a. We obtained the ground state with
full spin polarization for $J\simeq U' (\gtrsim 5t)$. 
The appearance of ferromagnetism in the $J<U'$ region can be well
understood with the effective Hamiltonian (\ref{eq:H_eff}). 
This ferromagnetic ground state has a strong alternating correlation 
in the orbital degrees of freedom. This is also consistent with the
argument from the effective Hamiltonian. In the perfect 
ferromagnetic ground 
state, the orbital degrees of freedom have isotropic (Heisenberg)
antiferromagnetic interaction with pseudo-spin and hence the alternating
correlation decays in a power form. 
The phase boundary for $J<U'$ approaches an asymptote $J=\alpha U'$
with $\alpha \simeq 0.35$ for large $U'$. 
The asymptote corresponds to the ground-state 
phase boundary of the effective Hamiltonian.\upcite{Sakamoto-MK}
The paramagnetic state for $0\lesssim J < \alpha U'$ may have 
the same properties as the ground state of the $SU(4)$ 
model at $J=0$.\rcite{Sutherland}{Yamashita-SU}
The ferromagnetic phase extends to the parameter region $J>U'$, as well. 
Though this region $J>U'$ is unrealistic, it is of
interest from the viewpoint of triplet superconductivity. 
\marker{An attractive}
force \markers{acts} between electrons with parallel spins due
to the Hund coupling and 
the present model might have some relevance to experimental 
results \marker{on} organic superconductors.\upcite{Lee-NDC} 
For $J>U'\gg t$ two electrons are 
paired to form a hard-core boson with spin unity.
The perturbation
due to the hopping term in eq.\ (1) leads to the effective 
Hamiltonian for these bosons, which includes hopping, 
repulsion and antiferromagnetic spin interaction between
nearest neighbor bosons. The effective Hamiltonian does not favor 
ferromagnetism and indeed the ferromagnetic phase does not \marker{extend}
to a 
region with large $J-U'$ in our numerical calculations. The slope of
the phase boundary approaches unity for large $U'$. 
We note that only the systems with $N_{\rm e}$ = even are
used in determining the phase diagram. In fact we found a large
difference in the phase boundaries for $J>U'$ between systems with even
$N_{\rm e}$ and odd $N_{\rm e}$. We consider that the results for
small odd $N_{\rm e}$ are strongly affected by the existence of an
unpaired electron and \marker{are} not useful \marker{for extracting}
bulk properties. 

Next we show the ground state phase diagram for 
$n=0.5$ in Fig.\ 1b. The ferromagnetic phase 
expands compared to that in the quarter-filled case both for $J>U'$
and $J<U'$, and the size dependence is very weak. 
It is remarkable that ferromagnetism is realized for
rather weak Hund coupling, that is, $J\simeq 2t$ for $U' \simeq 5$. 
Recently Hirsch\upcite{Hirsch} argued that the Hund coupling is not
effective enough to realize ferromagnetism in systems with low
density ($n<1$) and \marker{that} ferromagnetic exchange interactions between
different sites are necessary to explain ferromagnetism in low density 
systems \marker{like} Ni. 
From the present results, we expect that a moderate Hund coupling
realizes ferromagnetism in a bulk system in one dimension. 
This behavior should be compared with the result for 
infinite dimensions where we could not find ferromagnetism for $n<1$. 
(See next section.) 
\begin{figure}[tbh]
  \begin{center}
    \epsfig{file=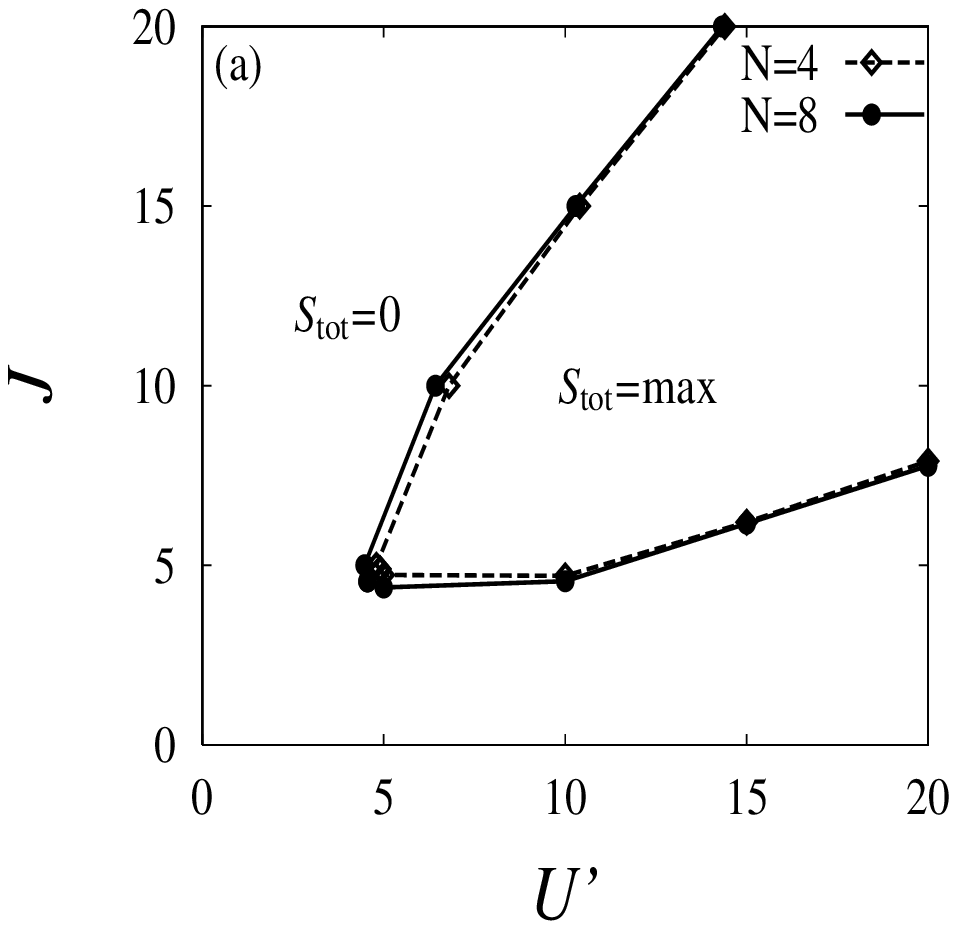,width=8.cm}\\
    \hspace*{-3mm}\epsfig{file=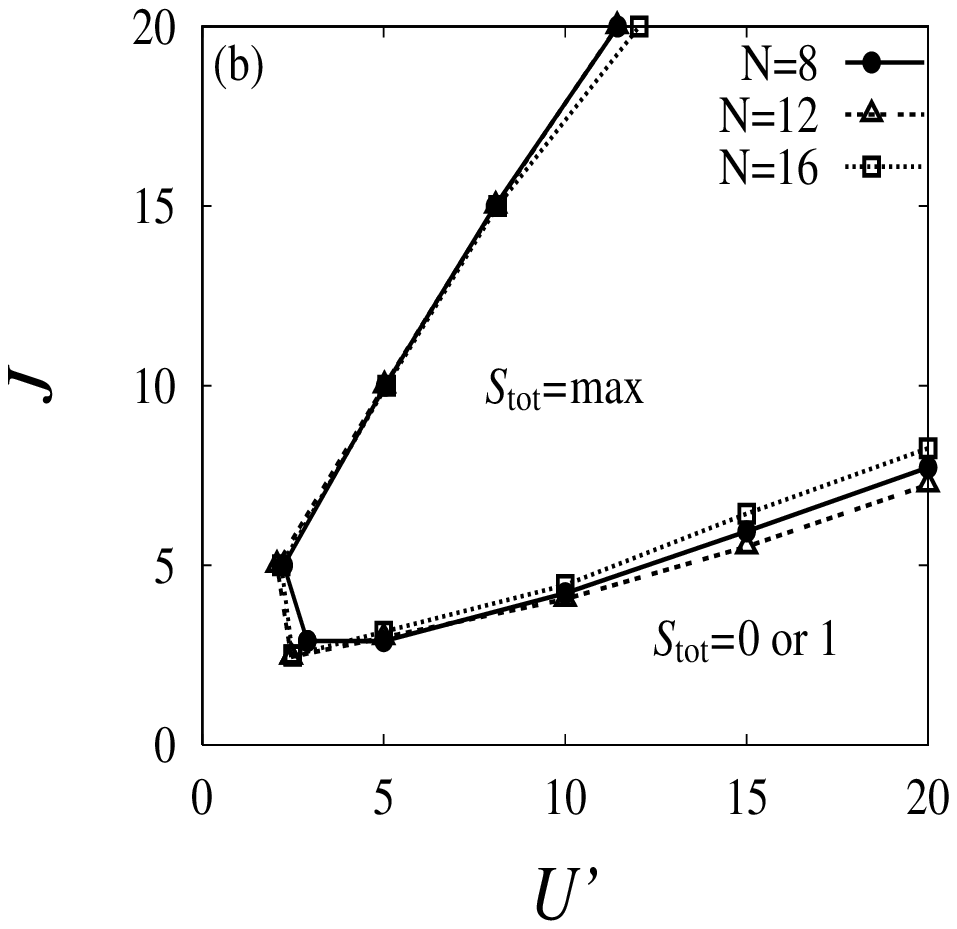,width=8.cm}\hspace*{-6mm}
    \epsfig{file=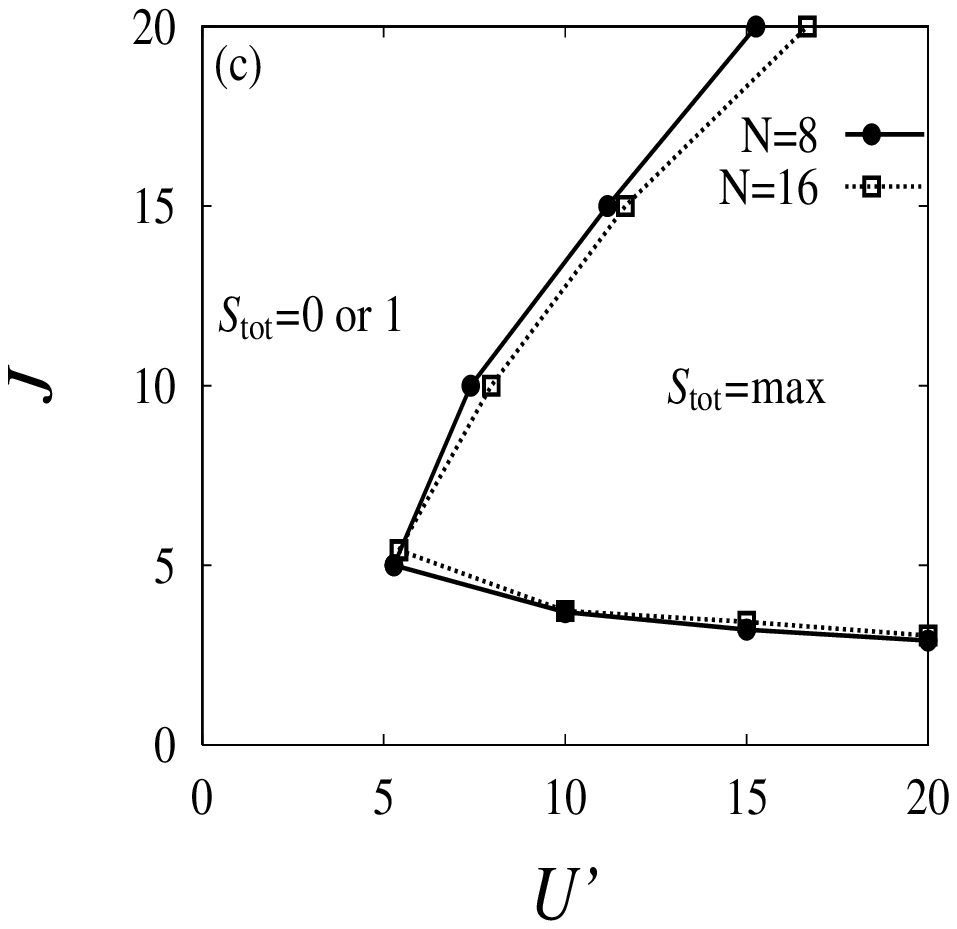,width=8.cm}\hspace*{-7mm}
  \end{center}
  \caption{Ground-state phase diagrams of the 1D doubly degenerate 
    Hubbard model for the filling $n=1$(a), 0.5(b) and 1.25(c). We assume $t$
    to be unity.  }
  \label{fig:PDDH1D}
\end{figure}

Finally, as an example of the case with $n>1$, we show the result for 
$n= 1.25$ in Fig.\ 1c. 
In this case 
ferromagnetism appears in a wider region than \marker{in} the
quarter filled case especially for small $J$. 
The lower phase boundary apparently approaches the line
$J=0$ for large $U'$. 
This enhanced stability of \marker{the} ferromagnetic state may be
understood as the result of \marker{the} \markers{``}double exchange
mechanism''. 
On the other hand the phase boundary for $J>U'$ is almost same \marker{as}
that for $n=1$. 

For all densities we found ferromagnetism on the line $J=U'$ with
strong $J(\gtrsim 5t)$. 
This result is consistent with the rigorous result in the limit
$J=U'\rightarrow\infty$ by Kusakabe and Aoki.\upcite{KusakabeA1}
We note that all the ferromagnetic ground states obtained above are
fully polarized. 
Since the ferromagnetic state \markers{is fully polarized} the spin 
 degrees of freedom are completely suppressed. 
The orbital degrees of freedom in the 
ground state \marker{are} mapped into the usual spin 
degrees of freedom in
 the single-band Hubbard model with the
interaction parameter $U'-J$.\upcite{Shen} 
Then we learn that, for $U'-J>0$, 
the pseudo-spin (orbital) correlation function decays with a power law 
as $\cos (|i-j|n\pi) \cdot |i-j|^{-\sigma}$. On the other hand, 
for $U'-J<0$, it shows an exponential decay but the pair-pair
correlation function of pseudospin-singlet (spin-triplet) pairs decays with a 
power law, which is a sign of quasi-long-range order of the triplet 
superconductivity.

\vspace{5mm}\noindent
{\bf Infinite Dimensional Model}

\vspace{5mm}
Next we discuss the model (\ref{2-band}) on a hypercubic lattice in
infinite dimensions.\upcite{Momoi-K} 
We scale the hopping integrals between nearest neighbor sites as
$t=\tilde{t}/2\sqrt{d}$ in $d$-dimensions and consider the limit $d =\infty$. 
Then the density of states (DOS) of each energy band has \marker{the}
Gaussian form 
$D(\varepsilon)=\exp(-\varepsilon^2/\tilde{t}^2)/\tilde{t}\sqrt{\pi}$.
We assume $\tilde{t}=1$ in the following. 
In this limit we can treat quantum fluctuations completely by taking 
local interactions into account and spatial correlations can be 
neglected.\upcite{GeorgesKKR}
The system is described in terms of a one-site effective action
which is determined self-consistently.
Generally one must rely on numerical methods to solve the effective action. 
In this study we 
approximated the action by that of a two-channel impurity model 
with finite ($n_s$) number of 
levels in each channel, and solved the impurity model by exact
diagonalization. \markers{The energy}
levels and mixing parameters of the impurity model \marker{were}
determined self-consistently. 
We searched for ground states which are uniform in space as well as 
those with two-sublattice structures.
Numerical calculations were done for 
$n_s = 5$ or 6. We mostly studied the system with $n_s=5$ and confirmed 
phase boundaries by using the system with $n_s=6$. We found that the 
results do not depend much on $n_s$.
We studied the ground state mainly at the fillings $n=1, 1.2$, and 0.8,
controlling the chemical potential. 

\begin{figure}[tbh]
  \begin{center}
  \begin{minipage}[t]{7.2cm}
    \begin{center}
      \epsfig{file=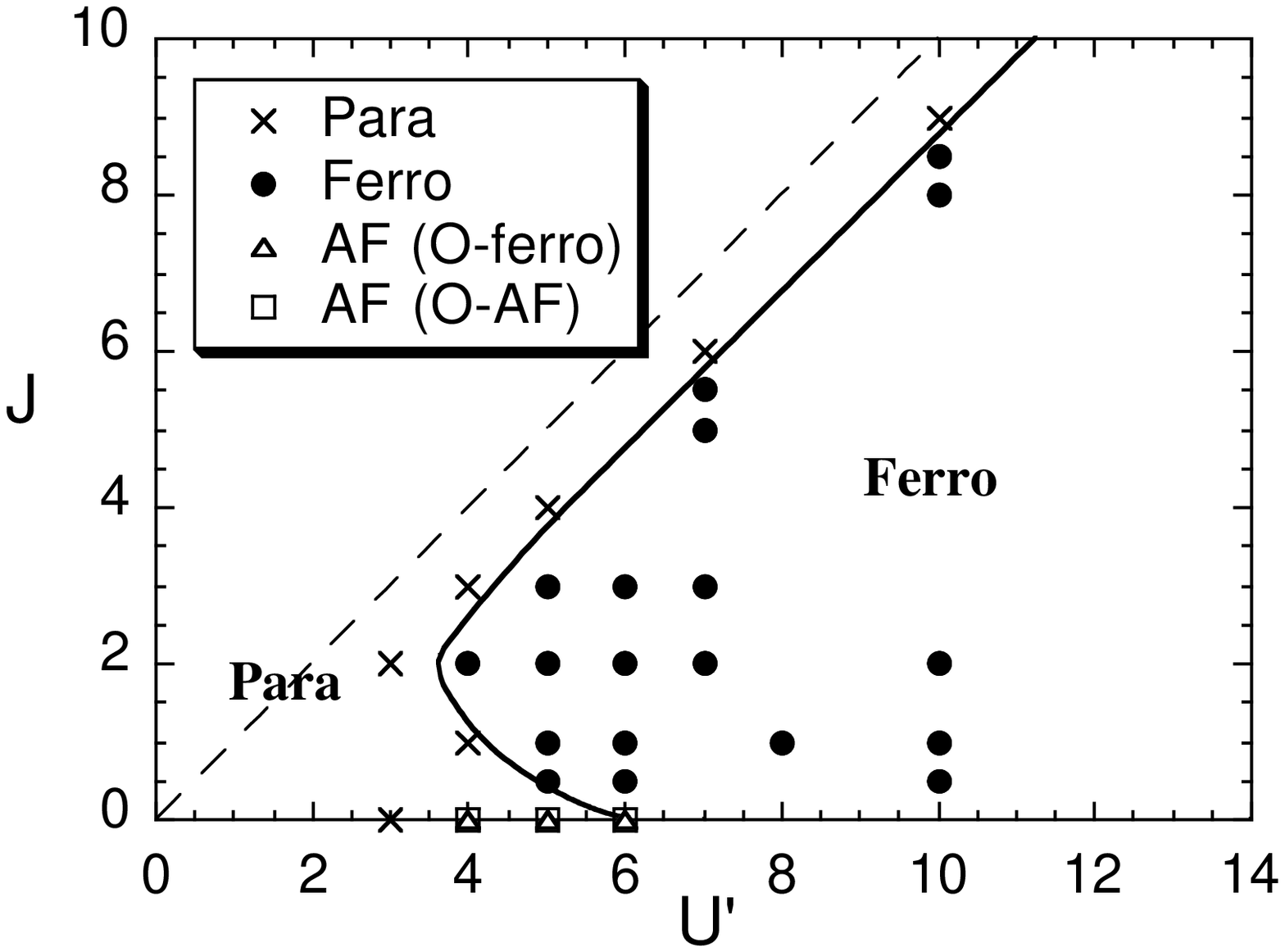,width=7.5cm}\\
      (a)
    \end{center}
  \end{minipage}
  \begin{minipage}[t]{7.2cm}
    \begin{center}
      \epsfig{file=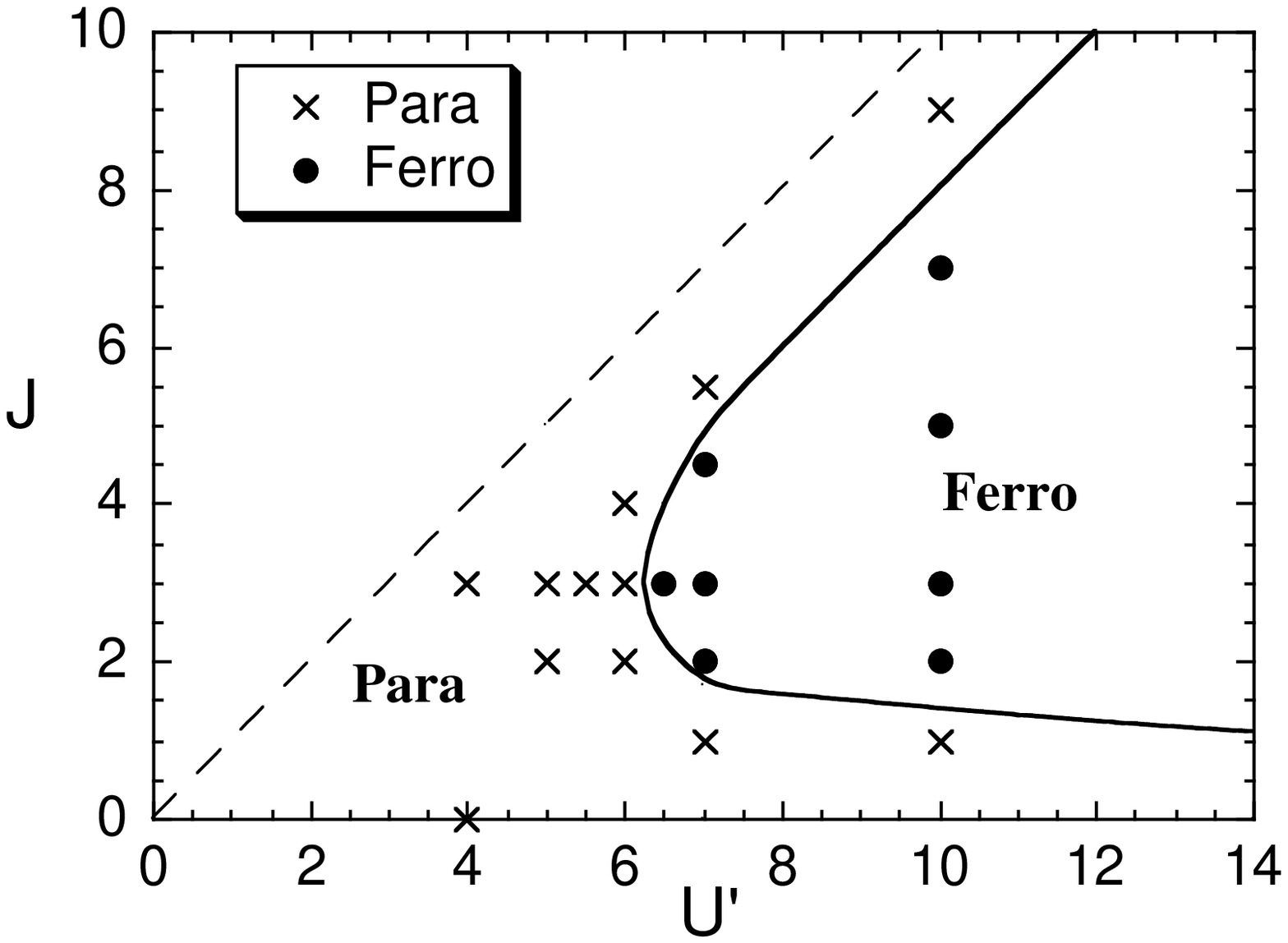,width=7.5cm}\\
      (b)
    \end{center}
  \end{minipage}    
  \end{center}
    \caption{Ground state phase diagram of the $D=\infty$ doubly degenerate 
      Hubbard model for $n=1$(a) and 1.2(b). } 
    \label{fig:PDDHinf}
\end{figure}
In the quarter-filling ($n=1$) case
we found paramagnetic and ferromagnetic ground states for $0<J<U'$ 
as is shown in Fig.\ 2a. 
Near the phase boundary two solutions coexist and
their energies cross over. We selected the ground state \marker{by}
comparing \marker{energies} and determined the phase diagram. 
The paramagnetic state \marker{obtained}
is spatially uniform and metallic. On
the other hand the ferromagnetic state has \marker{a} two-sublattice
structure with alternating orbital order, and is insulating. 
We found a narrow paramagnetic region for $J\simeq U'$.
Therefore the ferromagnetic phase seems to be confined within the
region $J<U'$, though we did not study the case with $J>U'$ in $d=\infty$.
 The ferromagnetic ground state 
appears even for $J\simeq 0$ for $U'\gtrsim 6t$. At $J=0$ the model 
possesses an $SU(4)$ symmetry. We found 
the coexistence of several ground states in this case. 

We show the phase diagram for $n=1.2$ in Fig.\ 2b. 
At this filling we \marker{obtain} a metallic ferromagnetic phase and a 
metallic paramagnetic one, both of which are spatially uniform.  
Both states have no orbital ordering. 
The area of the ferromagnetic phase is reduced in the phase diagram
compared to that of the insulating ferromagnetic phase at $n=1$.
As a hole-doped case, we studied the ground state for
$n=0.8$. At this filling we found only metallic ground states which are 
uniform in space and could not
find any magnetically ordered phase for $U^\prime \le 20t$.

\begin{figure}[tb]
  \begin{center}
    \epsfig{file=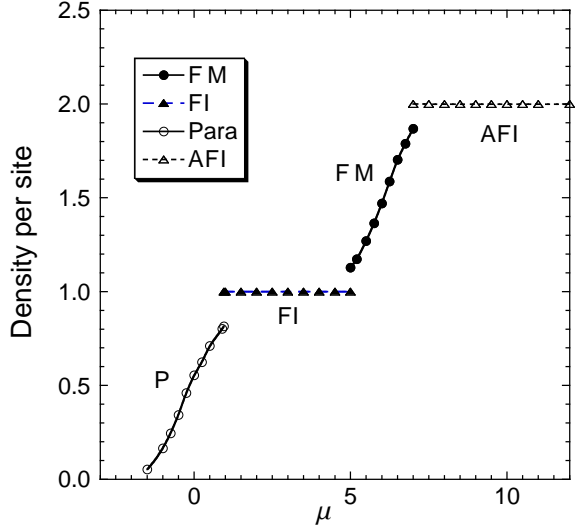,width=7.6cm}
  \end{center}
    \caption{The number density as a function of the chemical
      potential $\mu$ for $U'=10$ and $J=4$. The flat parts indicate
      insulating states.} 
\label{fig:density}
\end{figure}
We show the variation of the number density as a function of the chemical 
potential for $U'=10$ and $J=4$ in Fig.\ \ref{fig:density}. 
For \markers{these parameter values}
we have the paramagnetic metallic ground 
state for $0<n<0.82$. For $n=1$ the ground state is a ferromagnetic insulator.
\markers{The}
ferromagnetic metal is stable for $1.14<n<1.86$. The
antiferromagnetic insulator is realized for $n=2$. 
It is interesting that there are small jumps of $n$ \marker{on} both sides of 
quarter-filling. One is from $n=0.82$ to 1 and the other is $n=1$ to
1.14. There is another jump close to 
half-filling, i.e.\ between 
$n=1.86$ and 2. 
They imply that 
phase separation occurs for $n$ in these intervals.
Occurrence of 
phase separation was found also in the double exchange
model.\upcite{Yunoki-etal} 

\begin{figure}[tbh]
  \begin{center}
    \epsfig{file=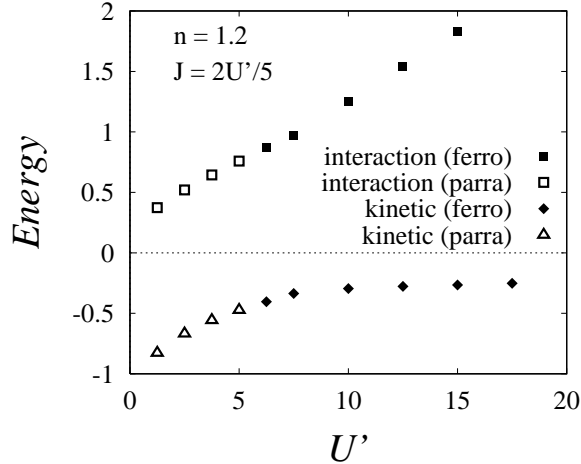,width=8.4cm}
  \leavevmode
  \end{center}
  \caption{The kinetic and the potential energy per site vs $U'$
    for $n=1.2$ on the line $J=0.4U'$.}
\label{fig:DDHenergy}
\end{figure}
We show the kinetic and interaction energy per site as a function of 
$U'$ on the line $J=0.4U'$ for $n=1.2$ in Fig.\ \ref{fig:DDHenergy}.
For these parameters the ground state is ferromagnetic for $U'\gtrsim 6$. 
The kinetic energy increases linearly with $U'$ for small $U'$ where
the ground state is paramagnetic. Then it starts to saturate 
and stays almost constant in the ferromagnetic phase. 
The interaction energy increases linearly with $U'$ but its slope
decreases as an effect of (local) correlations. The slope slightly
increases again in the ferromagnetic region and the potential energy
is nearly 
$(n-1)(U'-J)=0.12U'$ for large $U'$. \marker{The above}
result clearly shows that the
ferromagnetism is caused by 
reduction of kinetic energy \marker{rather than}
interaction energy. That means that the \markers{``}double
exchange mechanism" is the cause of 
ferromagnetism for this density. 

\marker{We} have seen \marker{above}
that the Hund coupling is effective both in one and 
infinite dimensions. Especially for $n>1$ the ferromagnetic ground state 
is realized in 
\marker{quite a} large parameter region in both dimensions. 
This result suggests 
that \marker{the}
``double exchange mechanism'' is quite effective in realizing 
the ferromagnetic state for $1<n<2$. We may 
expect that the situation is similar in two and three dimensions.
For $n=1$, 
ferromagnetism accompanied by \markers{alternating orbital} order 
is realized for $J<U'$. We found that the state is destabilized for weak $J$ 
in one dimension. This result may be understood as lower 
dimensionality \marker{stabilizing} the paramagnetic liquid state and
\marker{destabilizing} alternating orbital order. (In fact we have only 
quasi-long-range orbital order in one dimension.) The paramagnetic
state for $J=0$ is known to be an $SU(4)$ singlet
state.\rcite{Sutherland}{Yamashita-SU}

We found that the phase diagram is strongly dependent on the 
dimensionality for $n<1$. Though 
ferromagnetism is realized in a large parameter region in one dimension, 
we could not find it in infinite dimensions. \marker{Although}
our study \marker{does not exhaust the}
whole parameter region, it seems likely that there is no
ferromagnetism for $n<1$ in $d=\infty$. There may be a general
tendency \marker{for} ferromagnetism
in \marker{a} low density system \marker{to be} stabilized in one dimension.
In a 1D Hubbard model with nearly flat bands 
low density was found \marker{to be}
favorable for 
ferromagnetism.\upcite{Sakamoto-K} This tendency may
be understood as a 
result of the diverging DOS at the zone boundary in one dimension. 
Since we have quite different results for $d=1$ and $d=\infty$ for $n<1$,
\markers{a} 
study of the ground states in two and three dimensions \marker{is desirable} 
in order to answer the important question \markers{as to whether}
Hund coupling 
is effective in realizing ferromagnetism in low density systems \marker{like}
Ni.
\marker{It} should be noted that Ni has a fcc lattice structure, and 
its DOS has a sharp peak near the edge, which is similar to the
one-dimensional one. 

\vspace{15mm}\noindent
{\bf ELECTRONIC STATES IN THE DOUBLE EXCHANGE MODEL}

\vspace{5mm}
In this section we consider the so-called double exchange model (DEM),
which is composed of electrons in a single conduction band and
localized spins \marker{of magnitude} $S$ at all lattice sites. 
The electrons and localized spins interact through intraatomic Hund coupling.
The DEM may be \marker{the} simplest lattice model for electrons in manganites.
If we assume a single orbital instead of the doubly degenerate 
$e_{\rm g}$ orbitals and regard three electrons occupying 
$t_{\rm 2g}$ orbitals as a localized spin in Mn$^{3+}$ ions, we
obtain the DEM with $S=3/2$. We may consider also the DEM for
arbitrary $S$. For example, we may consider a level separation
$\Delta$ between two orbitals in the doubly degenerate Hubbard
model. If both $U'$ and $\Delta$ are much greater than $t$, then the
model reduces to the DEM with $S=1/2$ for $n>1$.
The DEM is described by the following Hamiltonian
\begin{eqnarray}
  \label{DEM}
  H=-t\sum_{\langle i,j \rangle\in {\rm n.n.},\sigma}
          (c_{i\sigma}^\dagger c_{j\sigma} + h.c.)
-J\sum_{i,\sigma, \sigma '} \mbox{\boldmath $S$}_{i}\cdot
   \mbox{\boldmath $s$}_{\sigma \sigma'}c_{i\sigma}^\dagger c_{j\sigma '},
\end{eqnarray}
where 
$\mbox{\boldmath $s$}={1\over 2}(\sigma^{x}, \sigma^{y}, \sigma^{z})$
and
$\sigma^{\alpha}$ denotes the Pauli matrix. 
The parameter $J$ in (\ref{DEM}) corresponds to $2J$ in (\ref{2-band}). 
Direct interactions between localized spins
 as well as the coupling between electronic and lattice degrees of
 freedom are neglected. For doped LaMnO$_3$ typical values 
of the conduction band width 
$2W$ and $J(2S+1)/2$ are thought to be $1\sim 2$ eV\upcite{Satpathy-PV}
and $2\sim 3$ eV, respectively. 
For large $JS$ the spin of an electron 
is always coupled parallel with the localized spin and forms a total spin
of size $S+1/2$. Since the original hopping term in (\ref{DEM})
conserves the spin of the electron, the hopping probability
between two neighboring sites is effectively reduced when localized
spins at these sites are not parallel to each other. The factor of the
reduction is given by $\cos \theta /2$ if the localized spins are
classical and make the angle $\theta$ between 
them.\upcite{Anderson-H} In the paramagnetic state the localized
spins are oriented randomly and as a result the conduction band is
narrowed due to 
the reduction of the hopping integrals as well as decoherence effects
\marker{due to scattering}. 
Band narrowing increases the kinetic energy of the paramagnetic state
and favors the ferromagnetic state.
Ferromagnetism due to this \markers{``}double exchange\markers{''}
mechanism was studied 
earlier.\upcite{deGennes,Kubo-O} 
The electronic states of the model with classical localized spins were
studied by use of dynamical mean 
field theory by Furukawa.\upcite{Furukawa}
Classical localized spins are not affected \marker{when they 
scatter} conduction electrons. \marker{Quantum mechanical}
spins may be flipped 
during the scattering processes. 
The effect of this \markers{``}spin exchange scattering\markers{''} 
was studied earlier 
\markers{using} the coherent potential
approximation (CPA).\rcite{Kubo72}{Takahashi-M}
The electronic states of the model with quantum spins in one and two
dimensions were recently studied by 
using numerical methods extensively.\upcite{Horsch-JM,Dagotto-etal}

\marker{The} CPA theory \marker{mentioned above}
treated a single electron in a system with randomly 
oriented localized spins and did not take into account 
the presence of 
other electrons. As a result the theory is valid only in the low density
limit, but in this limit it gives a qualitatively correct
description of the \marker{change in}
the electronic states due to interactions 
with localized spins. The single conduction band is modified by the 
interactions and 
the density of states splits into two bands for $JS\gtrsim W$. 
The lower band 
corresponds to 
electronic states with \marker{electron} spins parallel
to the localized spins (we call them 
``parallel electrons'') and
the upper band to those of ``antiparallel electrons''. 
The relative weights of the lower and the upper bands are
$(S+1)/(2S+1)$ and $S/(2S+1)$, corresponding to the total spin of a
site $S+1/2$ and $S-1/2$, respectively. 
If we naively consider these bands as a rigid one-particle density of states, 
the half-filled system (\mmarker{n=}
$N_{\rm e}/N=1$) cannot be an insulator since
the lower band is not full. 
Surely the half-filled system should be an insulator if $JS\gg W$.
We need a theory which realizes the insulating 
half-filled system for $JS\gg W$ in order to discuss the transport
properties of the DEM.

Let us first consider the atomic limit, i.e.\ the case with $t=0$.
The energy spectrum of the Green function in this limit 
is composed of four levels.
The lowest level at $\omega =-J(S+1)/2$ corresponds to the process
\marker{of creating}
a parallel electron at a site which is already occupied by an 
antiparallel one. The second one at $-JS/2$ comes from creating
a parallel electron at an empty site. The levels at $JS/2$ 
and $J(S+1)/2$ correspond to creating an antiparallel electron at an
occupied and unoccupied site, respectively. 
The spectral weights of these four levels are 
$(Sn-2 \langle \mbox{\boldmath $S$}_{i}\cdot \mbox{\boldmath $s$}_{i} 
\rangle )/(2S+1)$, 
$[(S+1)(2-n)+2 \langle \mbox{\boldmath $S$}_{i}\cdot 
\mbox{\boldmath $s$}_{i}\rangle ]/(2S+1)$, 
$[(S+1)n+2 \langle \mbox{\boldmath $S$}_{i}\cdot 
\mbox{\boldmath $s$}_{i}\rangle ]/(2S+1)$ and 
$[S(2-n)-2\langle \mbox{\boldmath $S$}_{i}\cdot \mbox{\boldmath $s$}_{i} 
\rangle ]/(2S+1)$, respectively, 
where $\mbox{\boldmath $s$}_{i}$ denotes the electron spin operator at
the $i$-th site.
When the hopping term is turned on, these discrete levels will 
broaden and compose four separate bands for $t\ll J$. They broaden as
$t$ increases and will finally merge into 
a single band for $t\gg J$.
We note that the first and 
third levels vanish in the low density 
limit ($n=0$) and that the weights of the two bands of the spectrum 
in the earlier one-electron CPA theory reproduce 
those of the second and fourth levels correctly.
As for the metal-insulator transition, 
we note that the two lower levels are fully occupied at $n=1$ in the 
exact atomic limit Green function.
Therefore, we can expect 
a Green function to reproduce the correct 
 insulating behavior at $n=1$ for $JS\gg W$, if it reduces to the exact
one in the atomic limit. 
We report \marker{below on the derivation of a} Green function in the one-site 
approximation which gives the correct atomic limit and \marker{also} reduces
to the 
one-electron CPA \marker{result}
in the low density limit. We calculate the resistivity 
$\rho$ in the framework of this approximation. 
\marker{A brief account of this work has already appeared and full details will
be published elsewhere.\upcite{rEdwardsGreenKubo}}

\begin{figure}[tbh]
  \begin{center}
  \leavevmode
    \epsfig{file=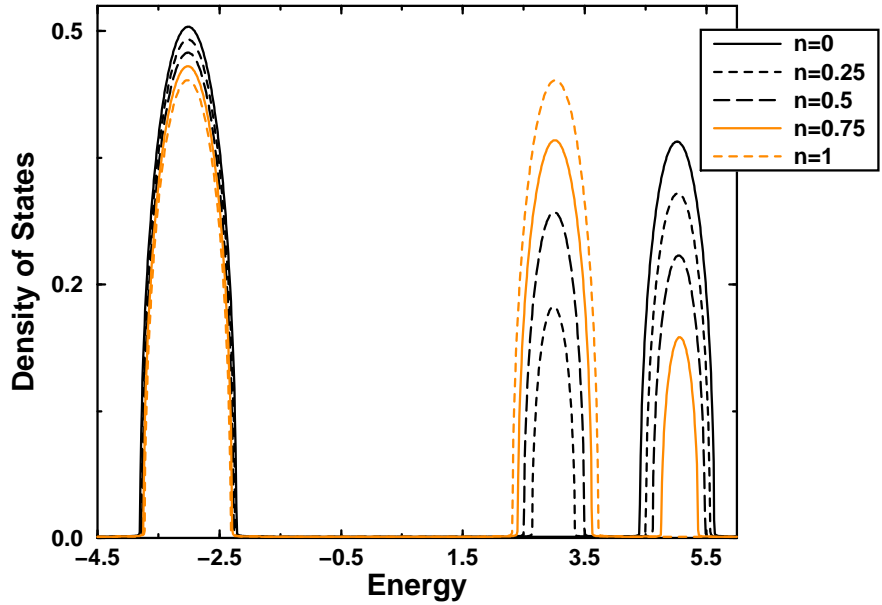,width=12.3cm}
  \end{center}
    \caption{The density of states obtained from the approximate Green
      function in the paramagnetic state for $J=4W$ and $S=3/2$ at the 
      filling $n=0$, $n=0.25$, \mmarker{n=0.5,} $n=0.75$ and $n=1.0$.}
\label{fig:DOS}
\end{figure}

In a single-site approximation the Green function $G_{\mbox{\boldmath
  $k$},\sigma} (\omega )$ is written as
\begin{eqnarray}
G_{\mbox{\boldmath $k$},\sigma} (\omega ) =
\{{\tilde G}_{\sigma}(\omega )^{-1} -(\epsilon_{\mbox{\boldmath $k$}} - 
J_{\sigma}(\omega ) ) \}^{-1},
\end{eqnarray} 
where
\begin{eqnarray}
J_{\sigma}(\omega ) = \omega - \Sigma_{\sigma}(\omega ) - {\tilde
 G}_{\sigma}(\omega )^{-1}. 
\end{eqnarray} 
Here $\epsilon_{\mbox{\boldmath $k$}}$ and $\Sigma_{\sigma}(\omega )$
denote the free band energy and the self energy, respectively.
The local Green function ${\tilde G}_{\sigma}(\omega )$ 
is related to the self-energy through the DOS of the free band energy 
$D_0 (\epsilon_{\mbox{\boldmath $k$}} )$ as 
\begin{eqnarray}
{\tilde G}_{\sigma}(\omega )= \int {{D_0 (x){\rm d}x}
\over{\omega - \Sigma_{\sigma}(\omega ) - x }}.
\end{eqnarray}
In order to close the above equations we need another equation for 
${\tilde G}_{\sigma}(\omega )$.
We \marker{follow}
 the equation of motion of the Green function along the
line of Hubbard.\upcite{Hubbard} 
We \marker{can}
close the equations for arbitrary $S$ in the paramagnetic
state. In the paramagnetic state we can omit the spin suffices and
the self-consistent equation for ${\tilde G}(\omega )$ is written as 
\begin{eqnarray}
{\tilde G}(\omega )= \sum_{\alpha\mathmarker{=\pm}}{
    {( E(\omega ) + \alpha J/2)(n/2)_\alpha + \alpha J/2
     \langle {\mbox{\boldmath $S$}}_{i}\cdot  {\mbox{\boldmath
         $s$}_{i}}\rangle  } 
\over{ \left(\mathmarker{E(\omega)}-\alpha\mathmarker{S}J/2\right)
\left(\mathmarker{E(\omega)}+\alpha J(S+1)/2\right) }}\ ,
\label{ParaGF}
\end{eqnarray}
where $E(\omega )\equiv \omega -J(\omega )$ and 
$(n/2)_\alpha\equiv \delta_{\alpha -} +\alpha n/2$.
\marker{The above}
set of equations \marker{reduces} to that of \markers{the}
one-particle CPA at $n=0$.
 We show in Fig.\ \ref{fig:DOS} the one-particle density of states 
for $J=4W$ and $S=3/2$ obtained
 from above set of equations by using the elliptic DOS
 given by 
\begin{eqnarray} 
D_0 (\epsilon )= 2/(\pi W^2 )\sqrt{W^2 - \epsilon ^2}.
\end{eqnarray}
 At $n=0$ the spectrum is composed of two bands centered at 
\mmarker{\omega = -JS/2} and $J(S+1)/2$, respectively. As $n$ increases 
the third band centered at $\omega =JS/2$ emerges and correspondingly 
the weight of the band at 
$\omega =J(S+1)/2$ decreases. There should be \markers{a} fourth band at
$\omega =-J(S+1)/2$ as well, but its weight is 
very small for $JS\gtrsim W$, since 
$\langle \mbox{\boldmath $S$}_{i}\cdot \mbox{\boldmath $s$}_{i}\rangle 
 \simeq  nS/2$ 
(In the calculation shown in Fig.\ \ref{fig:DOS} we approximated as 
$\langle \mbox{\boldmath $S$}_{i}\cdot \mbox{\boldmath $s$}_{i}\rangle 
=nS/2$ for simplicity).
The third band grows with $n$ and finally it takes over the second one
at $n=1$. 
At $n=1$ the lower two bands (the lowest band in Fig.\ \ref{fig:DOS})
are completely filled 
and the system becomes a Mott 
insulator at $T=0$ as is expected.
These three bands \marker{should}
be \marker{observable} by photoemission experiments \marker{on} manganites.
We note that the position of the band at $\omega \simeq JS/2$ will be
shifted to 
$\omega \simeq JS/2 + U$ in the presence of the intraatomic Coulomb 
repulsion $U$. For $J=\infty$ the lowest band has width 
\begin{eqnarray}
2{\bar W}=2W\sqrt{(S+1-n/2)/(2S+1)}
\end{eqnarray}
 for the elliptic DOS.
 The band narrowing factor is \marker{a} minimum at $n=1$
\marker{with the value} $1/\sqrt{2}$ independent of $S$.
It should be noted that 
this factor may \marker{depend considerably} on the choice of $D_0 $ as was
found for one-particle CPA.\upcite{Kubo72}

\begin{figure}[tbh]
  \begin{center}
  \leavevmode
      \epsfig{file=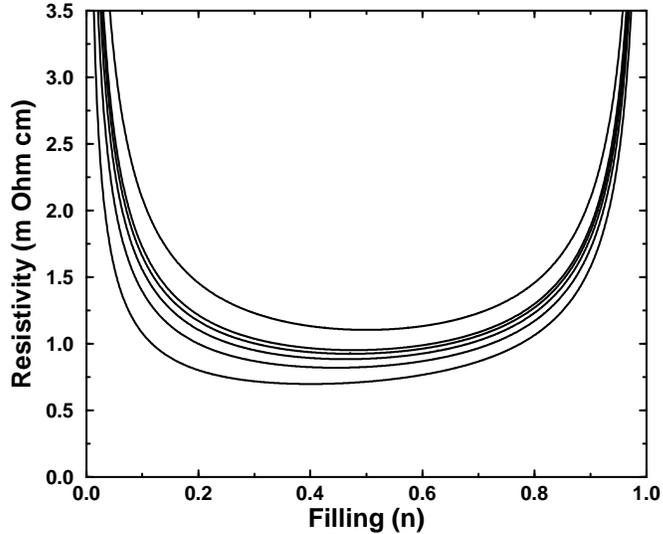,width=10cm}
  \end{center}
    \caption{Resistivity at $T=0$ (the paramagnetic state is assumed)
      is depicted versus $n$ for $J=\infty$. 
      Data for $S= 1/2 \sim 5/2$ and $\infty$ obtained by using 
      the elliptic DOS  are shown. Resistivity increases with 
      increasing $S$.}
\label{fig:rhoE}
\end{figure}
We calculate the DC resistivity $\rho$ by using the \marker{above} 
paramagnetic Green function in \marker{the} Kubo formula. Vertex corrections
do not enter in the calculation of the conductivity in the single-site 
approximation.
The expression of the static resistivity at $T=0$ is simplified 
by assuming the cubic tight-binding form of the hopping term in $H$.
\marker{We find}
\begin{eqnarray}\label{eres}
 \rho ^{-1} = 
%
{{2\pi e^2}\over{3a\hbar}}\int \epsilon{\rm d}\epsilon
 \phi(\epsilon )D_0 (\epsilon ),
\end{eqnarray}
where ${{\rm d}\phi(\epsilon )\over{{\rm d}\epsilon }} = 
A_{\mbox{\boldmath $k$}}(\mu )^2 |_{\epsilon_{\mbox{\boldmath $k$}}=\epsilon}$
and $A_{\mbox{\boldmath $k$}}(\mu )$ is the spectral weight function of the
 Green function $G_{\mbox{\boldmath $k$},\sigma} (\omega )$ at $\omega =\mu$.
\marker{This expression is evaluated using the elliptical approximation to the
DOS, both in $D_0(\epsilon)$ itself and in $\phi(\epsilon)$ via the Green
function calculated above. In Fig.\ \ref{fig:rhoE}
we} show the \marker{resistivity obtained} 
for $J=\infty$ \marker{as \marker{a function} of $n$ for} various values of 
$S$. We used 5{\AA} for the lattice constant $a$. 
 Note that the correct insulating behavior is obtained 
for $n=0$ and $n=1$. We \marker{find} that $\rho$ hardly
\marker{depends} on $J$ for $JS\gtrsim 5W$. \marker{The resistivity}
$\rho$ is of order of m$\Omega$cm for $0.1\lesssim
n\lesssim 0.9$ \marker{and this} is much smaller than
 typical experimental values 
for doped LaMnO$_3$ except for the case of La$_{1-x}$Sr$_x$MnO$_3$ 
with $x(=1-n)\simeq 0.3$.\upcite{Urushibara-etal}
The result shows also too weak \marker{a} dependence on $n$ compared to the
experimental result. 
Furukawa\upcite{Furukawa} calculated $\rho$ by using the Lorentzian DOS
for $S=\infty$ and obtained good \marker{agreement} with experimental result
at $x=0.2$. 
We also calculated $\rho$ by using the Lorentzian DOS defined as
\begin{eqnarray}
D_0 (\epsilon )= \mathmarker{W /[\pi (\epsilon ^2 + W^2 )]}
\end{eqnarray}
\begin{figure}[tbh]
  \begin{center}
  \leavevmode
    \epsfig{file=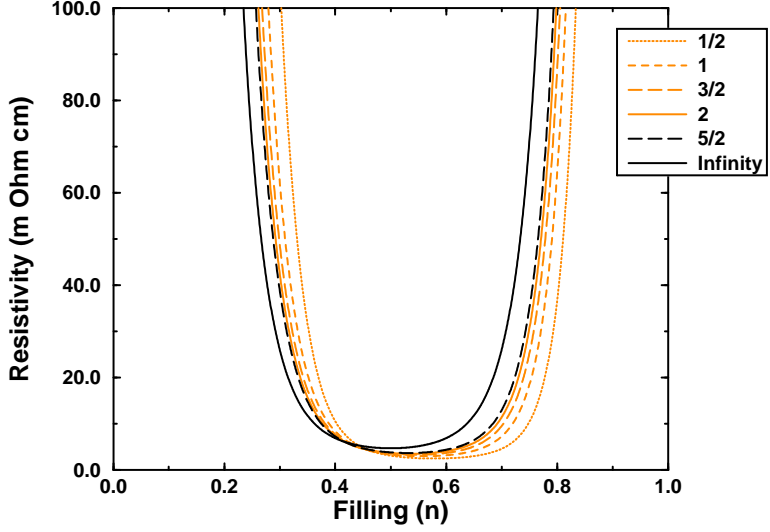,width=10.5cm}
  \end{center}
    \caption{Resistivity at $T=0$ for Lorentzian DOS.}
\label{fig:rhoL}
\end{figure}
\markers{to calculate $\phi(\epsilon)$ but retaining the elliptical
approximation to $D_0(\epsilon)$ so that the integral in eq.\ (\ref{eres})
converges.}
The results are shown Fig.\ \ref{fig:rhoL}. They 
show much stronger dependence on $n$ than those obtained for the
elliptic DOS, and increase very rapidly when the fermi level
approaches the band edge. The magnitude 
of $\rho$ at $x\simeq 0.2$ is several tens of m$\Omega$cm, which is
of the same order \marker{as the} experimental data.
However, the elliptic DOS is considered to be more realistic
than the Lorentzian one, since the Lorentzian DOS gives divergent
second moments. 
Hence the good agreement with experiment obtained for the
Lorentzian DOS may be an artifact. Our results suggest 
that \marker{scattering by} random 
localized spins \marker{is} not enough to explain the correct order of
magnitude of the resistivity in manganites. The present result is
obtained by treating dynamical aspects of the scatterings approximately. 
Effects of finite temperature and 
short-range correlations between
localized spins are also neglected, since our calculation has assumed
complete Fermi degeneracy and 
completely random configurations of localized spins. 
These effects may modify the above result to some extent, but we do
not expect that they will change the order of magnitude of the
resistivity. 
Therefore some other effects should be taken into account to explain 
experimental results.\upcite{Millis-LS,Roder-ZB,Millis-SM}

We also studied the magnetic \marker{properties}
of the system by using the Green
function. 
The self-consistent equation for the general magnetization for $S=1/2$
was obtained as 
\begin{eqnarray}
 &&{\tilde G}_{\uparrow}(\omega )= \\
 &&\sum_{\alpha =\pm}{
    {\langle n_{i\downarrow}^{\alpha}\rangle \left( E_{\uparrow}(\omega )
E_{\downarrow}^{\alpha}(\omega )-J^2 /8 \right) 
-(J/2)\left(\langle S^z_{i} n_{i\downarrow}^{\alpha}\rangle 
E_{\downarrow}^{-\alpha}(\omega )
-\alpha \langle S^{-}_{i} s^{+}_{i}\rangle
E_{\uparrow}^{-\alpha}(\omega ) \right) 
  }
\over{ E_{\uparrow}^{-\alpha}(\omega )
\left(E_{\uparrow}^{\alpha}(\omega )E_{\downarrow}^{\alpha}(\omega
  )-J^2 /4\right) 
 }},\nonumber
\end{eqnarray}
 where \mmarker{E_{\sigma}^{\alpha}(\omega ) = \omega -J_{\sigma}(\omega)
+\alpha J/4}.
We calculated the magnetic susceptibility $\chi $ for $J=\infty $ by
including a magnetic field and expanding $\langle S^z_{i} + s^z_{i}\rangle $ 
about the paramagnetic state.
We find that $\chi$ never diverges at a finite temperature for any
$0<n<1$, i.e.\ 
there is no ferromagnetic transition. At $n=0$ the correct Curie law 
\mmarker{\chi = (g\mu_{\rm B} )^2 {\tilde S}({\tilde S}+1)/(3k_{\rm B}T)} with
${\tilde S}=1/2$ was obtained. On the other hand \marker{for $n=1$ and
$J=\infty$ $\chi$ correctly} obeys the
Curie law with ${\tilde S}=1$ at high temperatures but it obeys the
law with \mmarker{{\tilde S}({\tilde S}+1)=2/15}
at low temperatures.

In the DEM with $J=\infty$ the ground state is proven to be
ferromagnetic in one dimension for any $0<n<1$.\upcite{Kubo} 
At present no reliable study of the ground state phase diagram of the
DEM with $S=1/2$
seems to be available in higher dimensions than one.
In three dimensions a high-temperature series expansion
analysis
suggests \marker{a} finite Curie temperature for all electron density between 0
and 1,\upcite{Roder-SZ} though the fully polarized state is not stable
for $0.12<n<0.45$.\upcite{Brunton-E} 
It is 
reasonable to expect that the ferromagnetic ground state
is stable 
in three dimensions in some density region. Therefore
we consider that our approximate Green function \marker{fails
to reproduce} the low temperature properties of the model \marker{correctly}.
It \marker{is} known \marker{that the analogous CPA in the Hubbard model}
does not give ferromagnetism
\marker{at any density\upcite{rFandE}}
or the correct Curie law at $n=1$.\upcite{Kawabata} 
 Our present approximation suffers from a similar 
failure and an improved treatment is necessary in order to discuss
the magnetic properties. We believe, however, that the present theory gives 
\marker{a}
qualitatively correct picture of the system in the paramagnetic state and
\marker{that} the
 paramagnetic resistivity obtained above is of the correct order of magnitude.
We need an
\marker{improved approximation to} the Green function to
study the magnetic 
properties of the system. A similar approach to that \marker{of}
Kawabata\upcite{Kawabata} for the strongly correlated Hubbard model
might be useful. 
 
\vspace{15mm}\noindent
{\bf SUMMARY}

\vspace{5mm} 
We discussed two topics on the Hund coupling in lattice systems
employing simplified 
models. First we examined the effectiveness of the Hund coupling in 
realizing ferromagnetism in the doubly degenerate Hubbard
model. In quarter-filled 
systems the insulating ferromagnetic state accompanied by alternating
orbital order 
was found stable. In more-than-quarter filling case 
metallic ferromagnetism is stabilized by \marker{the}
\markers{``}double exchange mechanism\markers{''}. \marker{The above}
results are common to one and infinite dimensions and we expect
\marker{them}
to hold in general dimensions. In less-than-quarter filled systems
the ferromagnetic ground state is stable in one dimension but not in
infinite dimensions. 
To study this case in two and three dimensions is an interesting
future problem. 

Secondly we examined the electronic states and the resistivity 
in the double exchange model by using the one-particle Green function. 
The splitting and narrowing of the one-particle spectrum due to the
Hund coupling 
were clarified in the framework of a single-site approximation. 
The resistivity due to the scattering by random localized spins \marker{was}
shown to be too small to explain the experimental results of doped manganites.
The present approximation failed to give the ferromagnetic state and
we need an improved treatment to study the properties at low temperatures.

\vspace{15mm}\noindent
{\bf ACKNOWLEDGMENTS}

\vspace{5mm}
We thank T.A.\ Kaplan, S.D.\ Mahanty, P.\ Horsch and N.\ Furukawa for useful 
discussions. K.\ K.\ and T.\ M.\ were supported by the JSPS Grant
No. 09640453. 
K.\ K.\ was supported by the \marker{EPSRC} Grant No.\ 
GR/L90804, \marker{and A.\ C.\ M.\ G.\ by an EPSRC studentship.}

\end{document}